\documentclass[a4paper, reprint, oneside, onecolumn]{revtex4-2}

\usepackage[english]{babel}
\usepackage[utf8]{inputenc}
\usepackage[labelfont=up]{subcaption}
\usepackage[x11names]{xcolor}
\usepackage{float}
\usepackage{mathtools}
\usepackage[top=1.in, bottom=1.in, left=0.5in, right=0.5in]{geometry}
\usepackage{amsthm, amssymb}
\usepackage{graphicx}
\usepackage{lipsum}
\usepackage{bm}
\usepackage[pdftex, pdftitle={Article}, pdfauthor={Author}]{hyperref}
\usepackage{wasysym}
\usepackage{enumitem}
\usepackage{appendix}
\hypersetup{colorlinks=true, citecolor=black, linkcolor=black, urlcolor=blue}

\let\vec\bm

\DeclarePairedDelimiter\ton{(}{)}
\DeclarePairedDelimiter\qua{[}{]}

\DeclareMathOperator{\me}{e}

\begin{document}

 \title{Time-Dependent Urn Models reproduce the full spectrum of novelties discovery}
 \author{Alessandro Bellina$^{1, 2, 3}$}
 \author{Giordano De Marzo$^{1, 2, 3, 4, 5}$}
 \author{Vittorio Loreto$^{3, 1, 2, 5}$}
  
\affiliation{
$^1$ Dipartimento di Fisica Universit\`a ``Sapienza”, P.le A. Moro, 2, I-00185 Rome, Italy.\\
$^2$ Centro Ricerche Enrico Fermi, Piazza del Viminale, 1, I-00184 Rome, Italy.\\
$^3$ Sony Computer Science Laboratories - Rome, Joint Initiative CREF-SONY, Centro Ricerche Enrico Fermi, Via Panisperna 89/A, 00184, Rome, Italy\\
$^4$ Sapienza School for Advanced Studies, ``Sapienza'', P.le A. Moro, 2, I-00185 Rome, Italy.\\
$^5$ Complexity Science Hub Vienna, Josefstaedter Strasse 39, 1080, Vienna, Austria.
}
  
  \date{\today} 
  
  \begin{abstract}
 
Systems driven by innovation, a pivotal force in human society, present various intriguing statistical regularities, from the Heaps' law to logarithmic scaling or somewhat different patterns for the innovation rates. The Urn Model with Triggering (UMT) has been instrumental in modelling these innovation dynamics. Yet, a generalisation is needed to capture the richer empirical phenomenology. Here, we introduce a Time-dependent Urn Model with Triggering (TUMT), a generalisation of the UMT that crucially integrates time-dependent parameters for reinforcement and triggering to offer a broader framework for modelling innovation in non-stationary systems. Through analytical computation and numerical simulations, we show that the TUMT reconciles various behaviours observed in a broad spectrum of systems, from patenting activity to the analysis of gene mutations. We highlight how the TUMT features a ``critical'' region where both Heaps' and Zipf's laws coexist, for which we compute the exponents.
\end{abstract}
\maketitle
\section{introduction}
Since the pioneering work of P\'olya~\cite{polya1930quelques}, urn models have received increasing interest in the scientific community~\cite{mahmoud2008Polya}. These stochastic models have been exploited in a broad range of disciplines and applied to several phenomena, such as genetic, epidemic contagion, voting processes, image segmentation, earthquakes, information filtering~\cite{hoppe1984polya, hayhoe2017polya, hayhoe2019curing, berg1985paradox, banerjee1999image, marsili1998self, marcaccioli2019polya}. More recently, urn models have been adopted to describe innovation processes and the occurrence of novelties. Inspired by Kauffman's concept of \textit{adjacent possible}~\cite{kauffman1996investigations}, Tria et al. introduced the so-called \textit{Urn Model with Triggering} (UMT)~\cite{tria2014dynamics}, whose key ingredient is the introduction of a conditional expansion of the space of colours within the framework of a classic P\'olya urn. By adjacent possible, we mean all those concepts, ideas, and objects we have not explored yet but whose exploration is one step away from being experienced. The idea behind the UMT is thus that whenever we experience something previously contained in the adjacent possible, i.e., we draw a ball whose colour was never drawn before, this triggers the expansion of the adjacent possible itself, a process schematised with the introduction of a given number of brand-new colours into the urn. The UMT reproduces both Heaps', Taylor's and Zipf's laws~\cite{tria2014dynamics, tria2018zipf, tria2020taylor}, often considered as the footprint of complexity~\cite{de2021dynamical}. In particular, the Heaps' law is related to the pace with which novelties occur, a proxy for the innovation rate. By denoting by $D$ the number of distinct elements observed in a system and by $t$ the total number of elements ever observed, Heaps' law reads as:
  \begin{equation}
 D(t)\sim t^{\gamma},
 \label{eq:Heaps}
  \end{equation}
\noindent with $\gamma\leq1$ being called Heaps' exponent. For instance, if we have a book in mind, $t$ would be the number of words composing it, while $D$ would correspond to the number of different words it contains. Heaps' law is observed in several systems ranging from language and cities to cryptocurrencies, and Wikipedia pages~\cite{tria2014dynamics,de2021dynamical, simini2019testing, marzo2022modeling}, and it is strictly connected with the presence of underlying power law distributions~\cite{de2021dynamical, lu2010zipf}. 
  
After its introduction, the UMT has been used to model, among others, innovation in the cryptocurrency market~\cite{marzo2022modeling}, interacting discovery processes~\cite{iacopini2020interacting}, and the emergence and evolution of social networks~\cite{ubaldi2021emergence}. However, despite its versatility, the UMT cannot, in its simplest form, only reproduce innovation processes described by functional forms different from the Heaps' law Eq.~\eqref{eq:Heaps}, as more generally observed in many real systems~\cite{hoppe1984polya, youn2015invention}. This limitation derives from the fact that the triggering and reinforcement processes of the UMT are time-independent. Whenever one draws a novel colour, one always introduces the same amount of new colours in the urn and reinforces them with the same amount of copies. Such an assumption may appear too strong if one thinks, for instance, of innovation in technological systems. In technological innovation, in fact, one can imagine that each new invention can potentially be recombined with what already exists to create further innovations. For example, the smartphone can be seen as the combination of a personal digital assistant, a phone, a camera, a music player, and a capacitive touchscreen. One may thus imagine scenarios in which the rate of expansion of the adjacent possible, though conditioned by the occurrence of novel events, instead of being constant, could be growing with the number of items previously introduced in a combinatorial fashion. This combinatorial expansion of the possibilities has been observed in patent data~\cite{youn2015invention}. Remarkably, the idea of a combinatorial adjacent possible is also behind the TAP (Theory of Adjacent Possible) equation proposed by Kauffman and collaborators~\cite{cortes2022tap, koppl2021explaining} to explain the abrupt economic transition our society experienced during the industrial revolution. 
  
This paper addresses the above-mentioned issue by introducing the Time-dependent Urn Model with Triggering (TUMT). In our generalisation of the UMT, we allow both the reinforcement and the triggering to be non-negative and monotonic sequences of $t$ and $D$, respectively. We do so by further elaborating on the previous works on time-dependent P\'olya's urns~\cite{pemantle1990time, pemantle2007survey, sidorova2018time}, where the effect of a reinforcement parameter that evolves in time was already considered. We derive explicit expressions for $D(t)$ depending on the functional form of the triggering and the reinforcement, and we show the model to be characterised by a very rich phenomenology, including a critical region, and by the presence of several distinct regimes. 

\section{The Urn Model with Triggering}
\label{definition_UMT}
First, let us describe the original UMT~\cite{tria2014dynamics}. One starts with an urn, initially containing $N_0$ balls of distinct colours. At each time step, a ball is randomly extracted from the urn, its colour is recorded and put back in the urn along with $\rho$ copies of it. Moreover, if the ball's colour is new, i.e., never drawn before, $\nu+1$ balls of distinct and brand-new colours are added to the urn. Here $\rho$ is the reinforcement parameter, while $\nu$ is the triggering one and sets the growth of the adjacent possible. 
The number of distinct colours appearing in the sequence of extractions $D$ grows with the length of the sequence $t$ as:
\[
  D(t)\sim
  \begin{cases}
  t \ \text{if}\ \nu>\rho\\
  t^{\frac{\nu}{\rho}} \ \text{if}\ \nu<\rho
  \end{cases}
\]
\noindent, and the model thus shows Heaps' law. The UMT has also been shown to reproduce Zipf's law with exponent $\rho/\nu$, meaning that the frequency $f(R)$ of the $R$-th most common colour ($R$ is rank) satisfies the following:
\[
  f(R)=\frac{f(1)}{R^{\frac{\rho}{\nu}}}.
\]
\noindent This result is the standard one would expect since Heaps' exponent $\gamma$ and Zipf's exponent $\alpha$ are known to be connected as $\gamma=1/\alpha$ for $\alpha<1$ and $\gamma=1$ otherwise~\cite{lu2010zipf, de2021dynamical}.
To generalize the UMT, we consider a slightly different definition that displays the same scaling exponent. In the original formulation of the UMT, when an old colour is extracted, the number of already seen colours increases by a term $\rho$. Conversely, if the extracted colour is new, the number of old colours increases by $\rho+1$, with the $+1$ deriving from the fact that the colour just drawn is no longer new. By denoting with $N_{new}^{(t)}$ the number of colours never drawn from the urn at time $t$, and with $N_{old}^{(t)}$ those already drawn, the characteristic matrix $\mathcal{M}$ of the process is given by: 
\[
 \mathcal{M}=
 \begin{pmatrix}
 \rho & 0\\
 \rho+1 & \nu.
   \end{pmatrix}
\]
\noindent from the characteristic matrix, one can write the evolution of the urn in vectorial form as:
\[
\begin{pmatrix}
 N_{new}^{(t+1)}\\
 N_{old}^{(t+1)}
   \end{pmatrix} =
   \begin{pmatrix}
 N_{new}^{(t)}\\
 N_{old}^{(t)}
   \end{pmatrix} + 
   \mathcal{M}\cdot \vec{\mathcal{E}},   
  \]
\noindent where $\vec{\mathcal{E}}$ is the event random vector and satisfies the following:
\begin{equation}
   \vec{\mathcal{E}}=
   \begin{cases}
 \begin{pmatrix}
  1\\
  0
 \end{pmatrix}
 \ \text{if an old element is extracted}\\
 \\
 \begin{pmatrix}
  0\\
  1
 \end{pmatrix}
 \ \text{if a new element is extracted}
   \end{cases}.
\end{equation}

\noindent We now modify the formulation of the UMT model by proposing instead a reinforcement of $\rho-1$ when a novel item is drawn from the urn. In this way, the reinforcement is the same whether reinforcing an old or a new ball. With the above change, the characteristic matrix now reads:
\begin{equation}
   \mathcal{M}=
   \begin{pmatrix}
 \rho & 0\\
 \rho & \nu
   \end{pmatrix}.
\end{equation}
\noindent The alternative definition of the UMT leaves the statistical properties of the model unaltered while making it easier to analyse. For a detailed discussion of this variant of the UMT model, we refer the reader to Appendix~\ref{appendix_A}. 

\section{The Time-dependent Urn Model with Triggering}
\label{definition_TUMT}

After having slightly redefined the UMT, we are ready to introduce the Time-dependent Urn Model with Triggering (TUMT). We define it simply by letting the reinforcement and the triggering parameters be time-dependent:
\begin{equation}
\rho\to\rho_t \;\; \nu\to\nu_D.
\end{equation}
\noindent The new characteristic matrix of the process now reads: 
\begin{equation}
   \mathcal{M}=
   \begin{pmatrix}
 \rho_t & 0\\
 \rho_t & \nu_D
   \end{pmatrix}
   \label{eq:matrix_TUMT}
\end{equation}
 \noindent with $\rho_t$ and $\nu_D$ being monotonic and non-negative sequences:
\begin{equation}
 \left\{
 \begin{array}{cc}
 \rho_t  \sim  t^{\phi}\\
 \nu_D  \sim D^{\psi}
 \end{array}
 \right.
\end{equation}
Note that while $\rho_t$ explicitly depends on $t$, i.e., on the number of draws, $\nu_D$ depends on $t$ trough $D$. As we mentioned in the introduction, the idea is that the expansion of the adjacent possible should depend on the number of available elements $D$, which can then be recombined to generate further possibilities. This structure aligns with what is observed empirically (e.g., in patents) and hypothesised in the TAP equations~\cite{cortes2022tap, koppl2021explaining}. 
It is important to remark that both $\rho_t$ and $\nu_D$ can generally be non-integer quantities and smaller than one, implying that the urn will asymptotically freeze and innovation would become harder and harder. The prescription to have non-integer quantities is to add to the urn $\lfloor \rho_t \rfloor$ balls with unitary weight (as in the standard UMT) and another ball with weight $\rho_t - \lfloor \rho_t \rfloor$. Here $\lfloor \rho_t \rfloor$ denotes the integer part of $\rho_t$, and the exact mechanism also applies to $\nu_D$. 

\noindent Further, we introduce the reinforcement and triggering time-series $P_t$ and $N_D$, as:
  \[
   \begin{cases}
 P_t=\sum_{i=1}^t\rho_i\\
 N_D=\sum_{i=1}^D\nu_i.
   \end{cases}
  \]
\noindent In the following, we shall show that the TUMT features three different regimes depending on the ratio between $\phi$ and $\psi$. 

\section{Pace of novelties}
We start our analysis of the TUMT by studying how the number of distinct elements grows as a function of the sequence length. We can write the number of new elements contained in the urn as: 
  \[
   N_{new} = N_0 + \sum_{i=1}^D\nu_i = N_0 + N_D.
  \]
\noindent The total number of elements in the urn, $\mathcal{N}$, is instead given by:
  \[
   \mathcal{N} = N_{new} + N_{old} = \sum_{i=1}^t\rho_i + N_0 + \sum_{i=1}^D\nu_i = N_0 + P_t + N_D.
  \]
\noindent The probability of drawing a new colour from the urn is given by $N_{new}/\mathcal{N}$, and thus $D_t$ evolves according to the following equation:
  \[
   D_{t+1} = D_{t} + \frac{N_{new}}{\mathcal{N}} = D_{t} + \frac{N_0 + N_D}{N_0 + P_t + N_D},
  \]
\noindent which, for large $t$, can be approximated as:
\begin{equation}
 \frac{d D(t)}{dt}=\frac{N_0 + N_D}{N_0 + P_t + N_D}.
 \label{eq:D(t)_general}
\end{equation}
Depending on the values of the exponents $\phi$ and $\psi$, describing the time dependence of the reinforcement and the triggering processes, respectively, this equation shows different behaviours that we investigate separately in the following subsections. 
   
\subsection{$\phi > \psi$, Reinforcement stronger than triggering}
Let us first consider the case $\phi > \psi$. In this case, the reinforcement process grows faster with time than the triggering process (i.e., adding new elements). For the moment, we suppose $\phi, \psi>-1$ such that the series can be approximated with the corresponding integrals $P_t\approx t^{a}$ and $N_D\approx D^{b}$, with $a=\phi+1$ and $b=\psi+1$. More generally, we can define $a = max(\phi+1, 0)$ and $b = max(\psi+1, 0)$, since when $\phi$ or $\psi$ are smaller than $-1$, the corresponding integrals converge to a constant. Substituting these expressions into Eq.~\eqref{eq:D(t)_general} one gets:
\[
 \frac{d D(t)}{dt}=\frac{N_0 + D^b}{N_0 + t^a + D^b}\approx \frac{D^b}{t^a},
\]
where we used the fact that $a>b$ and we neglected $N_0$. The solution of this equation is given by:
\begin{equation}
 D(t)\approx \qua*{\frac{b-1}{a-1}\ton*{t^{1-a}-1}+1}^{\frac{1}{1-b}}.
\label{eq:D_general_I0_ab>0}
\end{equation}
\noindent A general approach to solve Eq.~\eqref{eq:D(t)_general} in the case $\phi > \psi$ is given in Appendix \ref{appendix_B}. We have two possibilities depending on the value of $a$. 

\begin{figure}[t]
 \includegraphics[width=\linewidth]{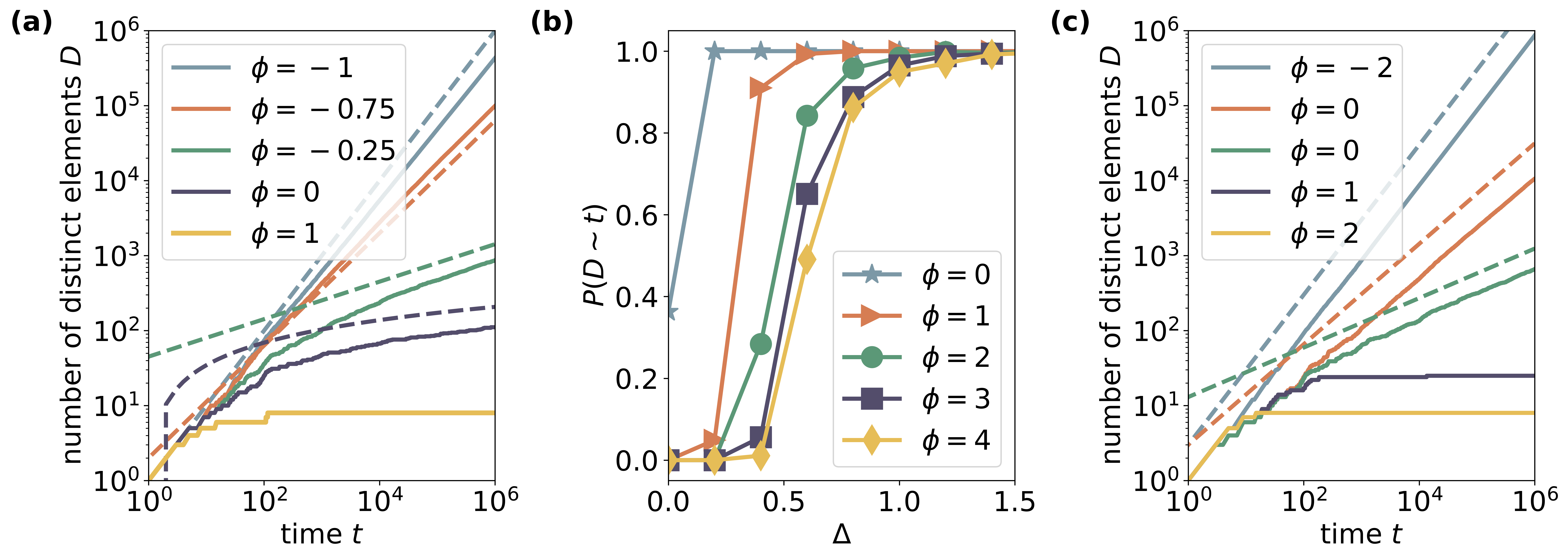}
 \caption{\textbf{Phenomenology of the model in the three regimes of the TUMT. (a)} Number of distinct elements $D(t)$ as a function of time $t$, in the regime $\phi> \psi$, with $\psi=-3$. When the exponent of the reinforcement sequence $\phi$ is smaller than $0$, we observe a power law behaviour for $D(t)$, with an exponent varying from $1$ when $\phi \leq 1$ to values smaller than $1$ when $-1<\phi<0$. For $\phi = 0$, we have $D(t) \sim \log t$, while when $\phi > 0$, one generally observes a saturation. All simulations were performed up to $t = 10^6$ with an initial condition $N_0 = 10$. \textbf{(b)} Probability that $D(t) \sim t$ in the regime $\phi < \psi$. The sequence of reinforcement is equal to $\rho_t = t^{\phi}$, while $\nu_D = D^{\phi+\Delta}$. As $\Delta$ increases, the probability of having $D(t) \sim t$ tends to $1$. This happens only when $\phi \geq 0$, as pointed out in the main text. To compute the probability, $10^3$ simulations were performed up to time $t=10^4$. \textbf{(c)} Number of distinct elements $D(t)$ as a function of time $t$, in the regime $\phi = \psi$. For $\phi < 0$ values, we generally observe a linear behaviour, $D(t) = ct$, for some constant $c$. $\phi = 0$ is the case of the standard UMT, which is able to reproduce sublinear behaviours for $D(t)$. Here, we display two combinations of constant $\nu = 1,2$ and $\rho = 3$ corresponding to the case $\phi=\psi=0$. Finally, for $\phi > 0$, we generally observe a saturation for $D(t)$. All simulations are performed as in the case (a).}
\label{fig:heaps}
\end{figure}
\begin{itemize}
 \item $\mathbf{a<1} \; (-1 < \phi < 0)$\\
In this case, the term $t^{1-a}$ is diverging, and it is thus dominating, from which we can perform the following approximation:
 \[
  D(t)\approx \frac{b-1}{a-1}t^{\frac{1-a}{1-b}} \sim t^{\frac{\phi}{\psi}}.
\]
where the last implication holds When also $-1 < \psi < 0 \; (b<1)$.
From this last expression, we thus see that even if the adjacent possible is expanding at a slower pace with respect to the reinforcement, the system still exhibits Heaps' law. In Fig.~\ref{fig:heaps}a we report the behaviour of $D(t)$ for $\psi=-3$ and several values of $\phi$ ($\phi= -1, -0.75, -0.25$). In this case, since $\psi \le -1$, then $b = max(\psi+1, 0) = 0$, thus $D(t) \sim t^{1-a}$.
\item $\mathbf{a>1} \; (\phi > 0)$\\
Now, the term in $t$ is sub-leading, and therefore one has:
\begin{align*}
  D(t) = \left( \dfrac{a-b}{a-1}\right)^{\frac{1}{1-b}} \left[ 1 + \dfrac{b-1}{a-b}t^{1-a} \right]^{\frac{1}{1-b}} \to \text{const}.
\end{align*}
For $\psi=-3$, one has $b=0$ and:
\begin{align*}
  D(t) = \left( \dfrac{a}{a-1}\right) \left[ 1 -\dfrac{1}{a}t^{1-a} \right] \to \text{const}.
\end{align*}
Differently from the previous case, $D(t)$ converges to a constant with a power law scaling, meaning that the system saturates and novelties are asymptotically absent. This case is exemplified in Fig.~\ref{fig:heaps}a for $\psi=-3$ and $\phi=1$, i.e., for $a=2$.
\end{itemize}
The cases $a=1 \; (\phi=0)$ and $b=1 \; (\psi=0)$ must be considered separately. For $a=1 \; (\phi=0)$ one has:
\begin{align*}
 & \dfrac{dD(t)}{dt} \approx \dfrac{D^b}{t} \qquad \implies \qquad D(t) \sim \ton*{\log t}^{\frac{1}{1-b}}.
\end{align*}
This case is exemplified in Fig.~\ref{fig:heaps}a for $\psi=-3$ and $\phi=0$, i.e., for $a=1$, by the behaviour $D(t) \sim \ton*{\log t}$. For $b=1 \; (\psi=0)$, one has:
\begin{align*}
 & \dfrac{dD(t)}{dt} \approx \dfrac{D}{t^a} \qquad \implies \qquad D(t) \sim e^{\frac{1}{1-a}(t^{1-a}-1)} \ \text{for}\ b=1.
\end{align*}
Note that the latter saturates being $a>b =1$. 

\noindent Up to now, we assumed $\phi>-1$. We treated already the case $\phi>-1$ and $\psi<-1$ in Eq.~\eqref{eq:D_general_I0_ab>0} where one sets $b=0$. If both $\phi<-1$ and $\psi<-1$, the two integral quantities, $P_t$ and $N_D$, will converge. As a consequence, using Eq.~\eqref{eq:D(t)_general} one obtains:
\[
\frac{d D(t)}{dt}=\text{const.}\to D(t)\sim t\ \text{for}\ a=b=0\ (\phi \ \text{and } \ \psi \le-1, ).
\]
It is also possible to consider other types of scaling for $\rho_t$ and $\nu_D$ but no conceptual differences emerge with respect to the cases already considered. Here we give some examples. 
\begin{itemize}
   \item $\rho_t\sim 1/t$ and $\nu_D\sim 1/t^{1+\delta}$ with $\delta >0$;\\
   In this case $P_t\sim \log t$ and $N_D=\text{const}$. From these considerations, it follows:
   \[
   \dfrac{dD}{dt} \sim \dfrac{const}{\log t} \implies 
   		D(t)\sim\text{LogInt}(t)\approx \frac{t}{\log t},
   \]
   i.e., a sub-linear growth.
   \item $\rho_t\sim e^t$ and $\nu_D\sim t^{b-1}$;\\
   Now we have $P_t\sim \me^t$ and $N_D\sim t^b$, leading to:
   \[
   	\dfrac{dD}{dt} \sim \dfrac{D^b}{e^t} \implies D(t) = \left[\dfrac{e^{-t}}{b-1} -\dfrac{b-1-e}{e}\right]^{\frac{1}{1-b}} \implies D(t)\sim 1-\me^{-t},
   \]
   which corresponds to an exponential saturation.
\end{itemize}
We conclude this paragraph by comparing, in Fig.~\ref{fig:heaps}a, our theoretical predictions and the growth of novelties observed in numerical simulations of the TUMT model. As it is possible to see, our computations capture the correct scaling in all the cases considered. 
 
\subsection{$\phi < \psi$, Triggering stronger than reinforcement}

We now consider the regime with $\phi < \psi$, i.e., where the exponent of the reinforcement is smaller than that of the triggering process. We only consider the case in which $N_D$ does not tend to be a constant; otherwise, we would recover the same scaling $D(t)\sim t$ already discussed above. Under this assumption we can neglect $N_0$ in Eq.~\eqref{eq:D(t)_general} so that one has:
\[
 \frac{d D(t)}{dt}\approx\frac{N_D}{P_t + N_D}.
\]
It is easy to realise that such an equation always admits $D(t)=t$ as a solution. Indeed, in this case, one has:
\[
\frac{d D(t)}{dt}\approx \frac{N_t}{P_t+N_t}\approx\frac{N_t}{N_t}=1,
\]
where we used the fact that $N_D=N_t$ and $\lim_{t\to\infty}N_t/P_t=\infty$, so that we can neglect $P_t$ in the denominator. However, $D(t)=\text{const}$, i.e., a saturating regime, can also be a solution provided that $P_t$ goes at least linearly in $t$. In this case, in fact, by neglecting $N_D$ with respect to $P_t$, one has:
\[
 \frac{d D(t)}{dt}\approx\frac{N_0 + N_D}{P_t + N_D}\approx\frac{\text{const}}{P_t}.
\]
If we set $P_t\sim t^a$ with $a<1$, we get $D(t) \sim t^{1-a}$, which contradicts the hypothesis $D(t)=\text{const}$. Consequently, we deduce that saturation and linear scaling coexist for $P_t=\mathcal{O}(t)$ or faster, as it can also be seen considering the probabilities of the two events. We can write the probability of saturation $\mathcal{P}_{sat}$ as:
   \[
 \mathcal{P}_{sat}=\prod_{t=T}^{\infty}\ton*{1-\frac{N_0+N_D}{P_t + N_D}}\approx\prod_{t=T}^{\infty}\ton*{1-\frac{\text{const}}{P_t}},
   \]
where we denoted by $T$ the time from which no novel element is extracted, and we used the fact that the right-hand side of Eq.~\eqref{eq:D(t)_general} gives the probability of extracting a novel colour. As shown in Appendix \ref{appendix_C}, this infinite product is non-null provided that $P_t$ diverges at least as $\mathcal{O}(t)$, and so both saturation and linear growth are possible. Which of the two regimes is observed will depend on the stochastic fluctuations of the discrete dynamics. Fig.~\ref{fig:heaps}b confirms a good agreement between theoretical predictions and numerical simulations. 
   
\subsection{$\phi = \psi$, Reinforcement and triggering on the same ground}
For $\phi = \psi$, the exponents for the time behaviour of the reinforcement and triggering processes are identical and one has that both $P_i$ and $N_i$ scale as $i^{\phi+1}$ or $\phi=\psi \ne 0$ (where $i$ is a generic index, which would be $t$ for $P_t$ and $D$ for $N_D$). We observe that, in this regime, $\nu_i$ and $\rho_i$ only differ by a constant that we define as $\mathcal{I}=\lim_{i\to\infty}\frac{\nu_i}{\rho_i}$. $\mathcal{I}$ quantifies the asymptotic relative importance of the reinforcement and triggering processes. Under these circumstances, one has that $N_D=\mathcal{I}P_D$, where $P_D$ is $P_t$ computed for the index $t=D$. Asymptotically, one can rewrite Eq.~\eqref{eq:D(t)_general} as:
\begin{equation}
 \frac{d D(t)}{dt}\approx\frac{\mathcal{I}P_D}{P_t + \mathcal{I}P_D}.
 \label{eq:D(t)_finiteI}
   \end{equation}
We first test a linear solution $D(t)=c\cdot t$ under the hypothesis $P_t = t^a$. By substituting this ansatz, we get:
\[
 \mathcal{I}c^{a-1}=\mathcal{I}c^{a}+1
\]
which has a solution in $c$ only provided that $a<1$. The linear scaling is the only possibility for $a<1$, while for $a>1$ it is easy to show that the saturating solution is unique. Finally, we can consider a sublinear scaling $D(t)=c\cdot t^k$ with $k<1$. Along the same lines as above, we obtain the following equation:
\[
 ck\cdot t^{k-1} = \frac{\mathcal{I}c^a\cdot t^{ka}}{t^a + \mathcal{I}c^a\cdot t^{ka}}\approx\mathcal{I}c^a\cdot t^{(k-1)a},
\]
that can be solved only for $a=1$ ($\phi=\psi=0$) and thus lead to $k=\mathcal{I}$. Since $a=1$, this corresponds to the case in which both $\nu_D$ and $\rho_t$ are constant, i.e., to the standard UMT model, and so as expected, we recover the result $k=\nu/\rho$~\cite{tria2014dynamics}. More details and computations for other functional forms of $P_t$ can be found in Appendix~\ref{appendix_D}. We show in Fig.~\ref{fig:heaps}c a comparison between theory and numerical simulations, confirming the validity of our results.
   
\section{Size distribution}
The other relevant aspect of Urn Models is the size distribution they produce, i.e., how the abundance of the different colours extracted from the urn is distributed. It is well known that the simple two-colour P\'olya's urn leads to a Beta distribution~\cite{polya1930quelques}, while the UMT gives rise to power law distributions and Zipf's law, as mentioned above. Here, we generalise these results to the TUMT model. First, we define $n_i$ as the total number of times the $i$th colour has been extracted and whose distribution we want to compute. Working directly with these quantities would be hard since they explicitly depend on the specific times $t^i_k$ of extraction of the $i$th colour. To circumvent this difficulty, we first focus on the quantities $N_i(t)$, given by:
\[
   N_i(t)= \sum_{k=1}^{n_i(t)} \rho_{t^i_k}.
\]
Under these conditions, one can write:
\[
   N_i(t+1)=N_i(t)+\frac{N_i(t)}{N_0+P_t+N_{D}}\rho_t,
\]
since the fraction appearing in the second term gives the probability of extracting the $i$th colour. This expression is closed and leads to the following different equation for $N_i(t)$:
  \begin{equation}
   \dfrac{d N_i(t)}{dt} \approx \dfrac{N_i(t) \rho_t}{N_0+P_t+N_D}.
   \label{eq:evolution_N_i(t)}
  \end{equation}
Depending on the regime, this differential equation shows different behaviours. 
\begin{figure}[t]
 \includegraphics[width=\linewidth]{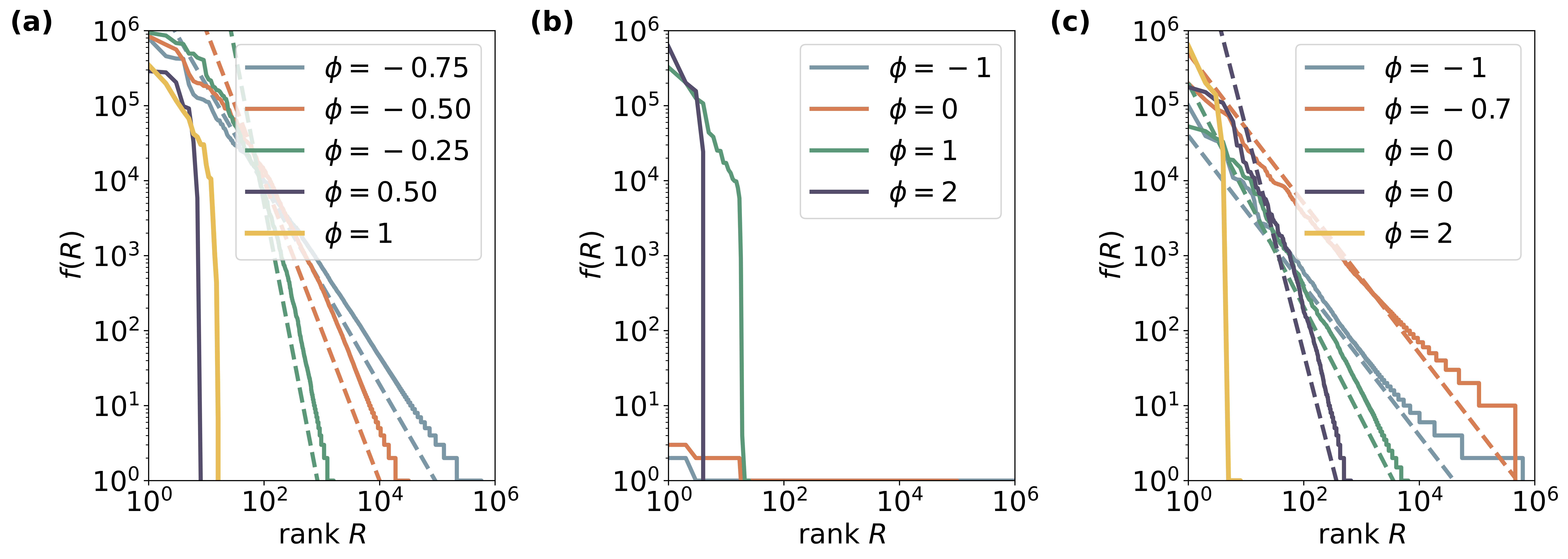}
 \caption{\textbf{Rank Size distributions in the three regimes of the TUMT. (a)} Rank-size plots in the case $\mathcal{I}=0$. When the exponent of the sequence of reinforcement $\rho_t = t^{-\phi}$ is between $-1$ and $0$, the distribution shows a power law tail. When instead $\phi > 0$, the saturation of $D(t)$ produces a step-like rank-size distribution, which can be seen as a power law with infinite exponent. \textbf{(b)} Rank-size plots in the regime $\mathcal{I}=\infty$. In this regime, we observe $D(t) \sim t$ with probability $1$ when $\phi < 0$, corresponding to a flat rank-size distribution. Alternatively, we can observe saturation with some probability for $\phi \geq 0$, leading to a step rank-size distribution. \textbf{(c)} Rank-size plots in the regime $0 < \mathcal{I} < \infty$. In this case, when $-1 < \phi < 0$, we generally have $D(t) = ct$, and this leads to a rank-size distribution with exponent $-1$. $\phi = 0$ corresponds to the standard UMT case. Here, we display two combinations of constant $\nu = 1,2$ and $\rho = 3$ corresponding to the case $\phi=\psi=0$. Finally, for $\phi > 0$, $D(t)$ saturates, leading to the usual step rank-size distribution. All simulations in the three cases are performed for $t=10^6$ with an initial condition $N_0 = 10$, except for the first three cases in (a), which are performed up to $t=10^7$ to decrease finite time effects.}
	  \label{fig:zipf}
   \end{figure}
\begin{itemize}

\item \textbf{Case $\phi>\psi$}\\
 When $\phi>\psi$ one can neglect the term $N_D$ as sub-leading asymptotically with respect to $P_t$ and approximate Eq.~\eqref{eq:evolution_N_i(t)} as:
 \[
  \dfrac{dN_i}{N_i} \approx \dfrac{\rho_t}{P_t}dt
 \]
Integrating both sides, and exploiting that $\frac{d}{dt} \log P_t = \frac{\rho_t}{P_t}$ one obtains: 
 \[
  N_i(t) = \rho_{t_i} \dfrac{P_t}{P_{t_i}},
 \]
where $t_i$ is the first time item $i$ is discovered, and we used the fact that the initial condition is $N_i(t_i)= \rho_{t_i}$. 

\item \textbf{Case $\phi<\psi$}\\
We first consider the situation in which $D \sim t$. In this case we can neglect the term $P_t$ in Eq.~\eqref{eq:evolution_N_i(t)} with respect to $N_t$ ($N_D$ in this case corresponds to $N_t$), obtaining in this way:
\[
  \dfrac{dN_i}{N_i} \approx \dfrac{\rho_t}{N_t}dt \implies \dfrac{N_i(t)}{\rho_{t_i}} \sim e^{\int_{t_i}^t \dfrac{\rho_t}{N_t} dt} = e^{F_t-F_{t_i}} = \dfrac{e^{F_t}}{e^{F_{t_i}}},
\]
where we introduced $F_t$ and $F_{t_i}$ as:
\[
  \int_{t_i}^t \dfrac{\rho_t}{N_t} dt = F_t - F_{t_i}
\]
Now, since $N_t$ is the integral of $\nu_t$ and $\nu_t > \rho_t$, the integrand $\rho_t/N_t$ is always smaller than $1/t$, which means the integral always converges for $t \to \infty$; as a consequence, $F_t \to const$ for $t \to \infty$. As an example, we can consider the case $\rho_t = t$, $N_t = t^3$, which yields:
\[
  N_i(t) \sim t_i \dfrac{e^{1/t_i}}{e^{1/t}}
\]
On the other hand, when $D$ saturates, it is straightforward to show that we obtain the same result of the case $\phi>\psi$, i.e., $N_i(t) \sim \rho_{t_i} P_t/P_{t_i}$.

\item \textbf{Case $\phi=\psi$}\\
In this case, Eq.~\eqref{eq:evolution_N_i(t)} becomes:
 \[
  \dfrac{dN_i}{N_i} \approx \dfrac{\rho_t}{P_t + \mathcal{I}P_D}dt
 \]
As already done when investigating the Heaps' law, we consider three possibilities. For $D \sim c\cdot t$ ($P_t = t^a$ with $0<a<1$), we obtain:
\[
 \dfrac{dN_i}{N_i} \approx \dfrac{a t^{a-1}}{t^a + \mathcal{I}(ct)^a}dt = \dfrac{a t^{-1}}{1 + \mathcal{I}c^a}dt \implies N_i(t) \sim t_i ^{a-1}\left(\dfrac{t}{t_i}\right)^{\dfrac{a}{1+\mathcal{I}c^a}}
\]
When $\rho_t = const$, i.e., when $\phi=0$, one has $P_t \sim t$ as in the standard UMT model. Finally, when $D$ saturates, one obtains the usual result $N_i(t) \sim \rho_{t_i} P_t/P_{t_i}$.
  \end{itemize}
In all the regimes considered above, from the expression of $N_i(t)$, it is straightforward to derive the cumulative distribution of the $N_i$. For example, when $N_i(t) = \rho_{t_i} P_t/P_{t_i}$, one can write:
\[
   p(N_i<N) = p\left(\rho_{t_i} \dfrac{P_t}{P_{t_i}} < N \right) = p\left(\dfrac{P_{t_i}}{\rho_{t_i}} > \dfrac{P_t}{N} \right) = 1- p\left(\dfrac{P_{t_i}}{\rho_{t_i}} < \dfrac{P_t}{N} \right)
\]
Let us focus on power-law like $P_t$, the situation for which it holds $\dfrac{P_{t_i}}{\rho_{t_i}} \sim t_i$. We have,
\[
   p\left(t_i < \dfrac{P_t}{N} \right) = p(N_i>N) = \dfrac{D \left(\dfrac{P_t}{N} \right)}{D(t)}
\]
However, we are interested in the distribution of the $n_i$, which represents the number of times the different elements have appeared in the sequence. In general, $n_i(t)$ can be simply related to the $N_i$ by considering the probability of drawing a ball of colour $i$ that reads $X_i(t)=N_i(t)/P_t$. In these terms, one has:
\[
\dfrac{d n_i}{dt} = X_i(t) = \rho_{t_i}/P_{t_i} = \dfrac{1}{t_i},
\]
where the last equality is exact only for power law-like $P_t$. Solving the differential equation, we obtain:
\[
   n_i(t) = \dfrac{t}{t_i}
\]
which gives, for the cumulative distribution: 
\[
   p(n_i>n) = p\left(t_i < \dfrac{t}{n} \right) = \dfrac{D \left(\dfrac{t}{n} \right)}{D(t)}. 
\]
In particular, whenever Heaps' law is observed $D \sim t^{\gamma}$, the expression above leads to a power law-like size distribution:
\[
   p(n_i>n)\sim \frac{1}{n^{\gamma}}
\]
and consequently, the system shows generalised Zipf's law with exponent $\alpha=1/\gamma$
\[
   f(R) = \frac{1}{R^{1/\gamma}}=\frac{1}{R^{\alpha}}.
\]
Here, $f(R)$ denotes the size of the $R$th largest element in the system, and $R$ is the rank. In Fig.~\ref{fig:zipf}, we compare the theoretical predictions and the rank-size plots observed in numerical simulations. The three panels correspond to the three different regimes we discussed above. More details on how to compute the rank-size distributions and the specific result for the cases reported in Fig.~\ref{fig:zipf} are reported in Appendix \ref{appendix_E}. It is worth pointing out that in the case $\phi=\psi$, whenever $-1 < \phi < 0$, one obtains Zipf's law with an exponent $\alpha=1$, a statistical regularity that has been observed in an astonishing number of systems ranging from natural language to cities~\cite{li2002zipf}. While several models have been proposed for explaining Zipf's law, the mechanisms underlying its universality are still to be fully understood. Our results show that Zipf's law is a very robust statistical property that, unlike other scaling exponents, can be found for a wide range of parameters for which innovation and reinforcement balance themselves (i.e., $\phi=\psi$).
\begin{figure}[t]
 \includegraphics[width=\linewidth]{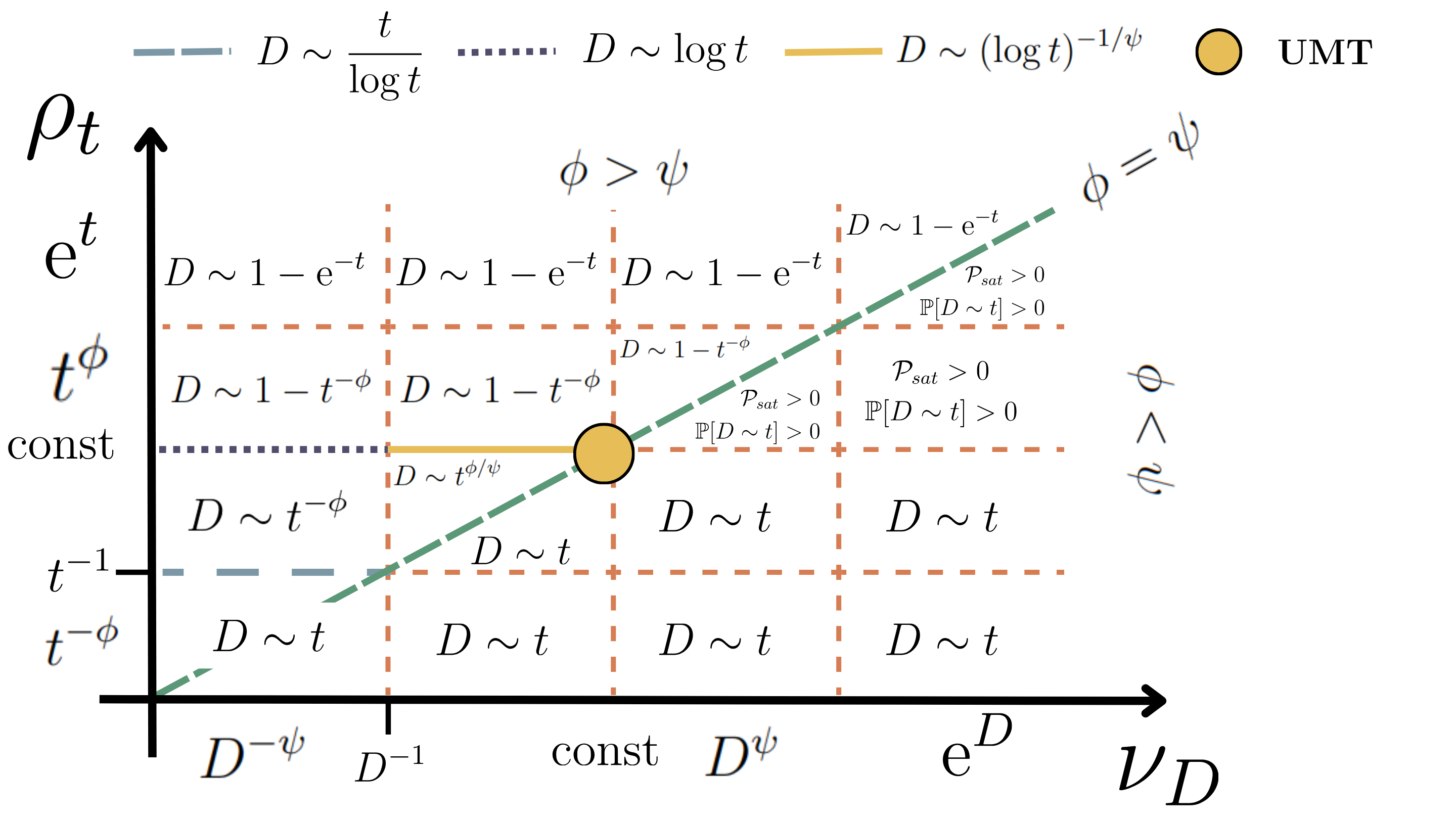}
 \caption{Graphical representation of the diverse phenomenologies displayed by the Time-dependent Urn Model with Triggering (TUMT). Each quadrant corresponds to a different choice of reinforcement and triggering sequences. The diagonal (green line) corresponds to the case $\phi =\psi$, and splits the plane in a region above the diagonal with $\phi > \psi$ and a region below the diagonal with $\phi < \psi$. Above the green line, one mostly observes saturation of $D(t)$. Still, for $-1 < \phi <0$, one observes a sublinear regime similar to that present in the standard UMT case when $\nu < \rho$. Below the diagonal, one generally observes $D \sim t$, but when $\phi > 0$, there is a finite probability for $D(t)$ to saturate. Along the diagonal, we can observe both $D \sim t$ ($\phi < 0$) and saturation ($\phi > 0$), while the yellow highlighted point $\phi = 0$ represents the standard UMT case.}
\label{fig:plane}
\end{figure}
\section{Conclusions}
Innovation and novelties are some of the driving forces of human society and individuals' lives. Despite the complex mechanisms that drive these processes, many innovation-driven systems are characterised by universal statistical regularities. For instance, the number of different patent code combinations and the number of cryptocurrencies evolves following a power law scaling and is thus described by Heaps' law. However, different behaviours, such as the logarithmic scaling in gene mutations or patent codes, have also been observed. Understanding how innovation dynamics works and how these statistical laws emerge is a timely issue that can help us pursue technological advancement. Still, the modelling of innovation processes is particularly challenging. One of the main challenges is the presence of an underlying space of items that enlarges as innovations occur. Kauffman's concept of Adjacent Possible provides a framework that helps understand how this expansion works. Urn models have been used to model innovation processes since the work of Hoppe on gene mutation, but only recently have they been integrated with the Adjacent Possible framework in the so-called Urn Model with Triggering~\cite{tria2014dynamics}. The UMT successfully reproduces the whole spectrum of Heaps' law, but one cannot reproduce other functional forms in its framework. Moreover, the UMT assumes a constant reinforcement of past choices and a constant rate for the expansion of the adjacent possible. While this is a reasonable approximation in some contexts, other systems may display an explicit dependence on time or their history.
  
Here, we introduced a generalisation of the UMT, the Time-dependent Urn Model with Triggering (TUMT), which solves the issues we mentioned above. As the UMT, the TUMT model is characterised by two parameters describing the reinforcement and the triggering processes. The former sets the tendency to exploit previous innovations, while the latter tunes the exploration of novelties and thus the expansion of the adjacent possible. By letting the reinforcement be a time-dependent quantity and making the triggering depend on the number of already realised innovations, we manage to unify several different innovation behaviours under the same framework. For instance, the combinatorial innovation processes underlining the TAP equation~\cite{cortes2022tap} and observed in patenting activity~\cite{youn2015invention} correspond to a triggering parameter that increases as the factorial of the number of previous innovations, as explicitly shown in Appendix \ref{appendix_F}. Depending on how the reinforcement evolves, this can give rise to a linear Heaps' law, as observed in the number of technological code combinations, or to a saturation regime, where the exploitation is too strong and innovation stops. Differently, when the exploitation is constant and the triggering parameter decreases over time, one observes a logarithmic growth of innovations. This behaviour is reminiscent of gene mutations modelled by Hoppe, whose phenomenology is reproduced by the TUMT; the analogy is explained in Appendix~\ref{appendix_F}. This also suggests that the logarithmic growth of the number of technological codes could be explained by an increasing difficulty in producing novel technologies, with innovation taking place thanks to the recombination of existing technological codes.

The TUMT shows a rich phenomenology governed by the interplay between the reinforcement and the triggering strengths, defined as the ratio between the series of the values for reinforcement and triggering and their specific functional forms. Fig.~\ref{fig:plane} shows a schematic summary of the phenomenologies possible in the framework of the TUMT model. It is worth pointing out the presence of a sublinear Heaps' scaling region in the $\phi>\psi$ regime (above the diagonal), where Zipf's law is also observed. Consequently, one observes the same critical behaviour of the UMT when the reinforcement and the triggering are not constant. As we already mentioned, another relevant region of the $\phi > \psi$ area is observed on the line of constant reinforcement. In this region, the TUMT presents the same logarithmic scaling observed in gene mutations and technological codes. The constant innovation rate line, in blue, is divided by the point corresponding to the UMT in two regions. Below the UMT, a linear Heaps' law is found; above it, the system saturates, and the number of innovations remains finite. Finally, when $\phi>\psi$, one generally observes a linear Heaps' law, even if a finite probability of saturation appears as the reinforcement becomes faster and faster.

By expanding the UMT framework to non-stationary systems, the TUMT widens the applicability of the adjacent possible idea. Still, it represents a strong simplification of how innovation takes place. One of the main ingredients neglected is the presence of an underlying structure that determines which innovations get triggered as the result of a previous breakthrough. For instance, we are not explicitly looking at innovation as the result of the combinations of previous technologies. From this perspective, leveraging a technology space, as introduced by~\cite{tacchella2020language}, would be very interesting to get a more realistic picture of innovation dynamics. Moreover, we are neglecting the possible presence of correlations, as those considered in the Urn Model with Semantic Triggering~\cite{tria2014dynamics}, and higher order interaction, as those investigated and modelled in~\cite{di2023dynamics}. Despite these limits, the TUMT successfully reproduces the statistical regularities observed in systems governed by innovation, starting from simple microscopic mechanisms. As such, it represents a valuable starting point for modelling novelties in non-stationary innovation-driven systems. 
  
%
\appendix

\section{Equivalence of the two definitions of the UMT}
\label{appendix_A}
In this Appendix, we show that the choice between the two characteristic matrices mentioned in Section \ref{definition_TUMT}
  \[   
  \text{(1)} \qquad \mathcal{M}=
  \begin{pmatrix}
   \rho & 0\\
   \rho + 1 & \nu
  \end{pmatrix} \qquad \text{or} \qquad \text{(2)} \qquad  
  \mathcal{M}=
  \begin{pmatrix}
   \rho & 0\\
   \rho & \nu
  \end{pmatrix}
  \]
does not significantly impact the phenomenology of the model. To establish this, we will derive the dynamic equations for the variable $D(t)$ to demonstrate that it follows Heaps' Law with consistent exponents. It is important to recall that the evolution of $D(t)$ can be expressed as:
\[
   D_{t+1} = D_t + p_{new}
\]
where $p_{new}$ represents the probability of selecting a colour that has never been observed. This probability can be computed as the ratio of the number of new balls to the total number of balls in the urn.

The number of new balls in the urn, regardless of the characteristic matrix $\mathcal{M}$ definition, is given by $N_0 + \nu D$, where $N_0$ represents the initial number of balls in the urn. The term $\nu D$ accounts for the fact that, with each triggering event, $\nu_1+1$ new colours are introduced, but the extracted ball becomes an old one, thus compensating for the $+1$.

  In contrast, the total number of balls is $N_0 + \rho t + (\nu+1) D$ in case (1). In this scenario, $\rho$ new balls are introduced at each time step, and during each triggering event, $\nu+1$ new balls of different colours are introduced.
  
  In case (2), the situation is slightly different. In non-triggering events, $\rho$ new balls are introduced, contributing to the term $(t-D_t)\rho$. $\nu+1$ new colour balls are introduced during triggering events, but only $\rho-1$ balls are of the extracted colour. This results in the two terms $(\nu+1)D_t+(\rho-1)D_t$. When considering the initial number of balls in the urn $N_0$, the total number becomes $N_0 + \rho t + \nu D_t$.

  Finally, we can express the evolution of $D_t$ and, by taking the continuous time limit, the evolution of $D(t)$ in both cases:
  \begin{align}
   \text{(1)} \quad & D_{t+1} = D_t + \frac{N_0 + \nu D_t}{N_0 + \rho t + (\nu+1)D_t} \quad \implies \quad \frac{dD(t)}{dt} \approx \frac{N_0 + \nu D(t)}{N_0 + \rho t + (\nu+1)D(t)} \nonumber \\
   \text{(2)} \quad & D_{t+1} = D_t + \frac{N_0 + \nu D_t}{N_0 + \rho t + \nu D_t} \quad \implies \quad \frac{dD(t)}{dt} \approx \frac{N_0 + \nu D(t)}{N_0 + \rho t + \nu D(t)} \nonumber
  \end{align}

  When considering the regime $\nu < \rho$, in both cases, the equations reduce to:
  \[
   \dfrac{dD(t)}{dt} \approx \dfrac{\nu D(t)}{\rho t}
  \]
  which leads to $D(t) = t^{\frac{\nu}{\rho}}$, recovering the sublinear Heaps' exponent.

  On the other hand, when $\nu > \rho$, the solution in both cases takes the form $D(t) = k t$, as evidenced by direct substitution into the equation for the evolution of $D(t)$. The constants in the two cases are:
  \[
   \text{(1)} \quad k = \dfrac{\nu-\rho}{\nu+1} \qquad \qquad
   \text{(2)} \quad k = \dfrac{\nu-\rho}{\nu} 
  \]

  Thus, in both cases, a Heaps' exponent of $1$ is recovered, with slightly different coefficients for $D(t)$.

 \section{Detailed computations in the case $\phi> \psi$}
 \label{appendix_B}

In this section, we will outline a general method for solving Eq.~\eqref{eq:D(t)_general} and provide a specific example of the procedure when $\phi> \psi$.

In this particular case, especially for increasing values of $P_t$, we can safely neglect $N_0$ and $N_D$ in the denominator of \eqref{eq:D(t)_general} relative to $P_t$. Furthermore, we can also disregard $N_0$ compared to $N_D$ in the numerator. This allows us to derive an approximate equation for the evolution of $D(t)$:
\[
   \dfrac{dD(t)}{dt} \approx \dfrac{N_D}{P_t}
\]
This equation can be solved using the separation of variables, regardless of the specific expressions for $P_t$ and $N_D$:
\[
   \dfrac{dD}{N_D} \approx \dfrac{dt}{P_t} \qquad \implies \qquad \int_{1}^{D} \dfrac{dD}{N_D} \approx \int_{1}^t \dfrac{dt}{P_t}
\]
Here, we establish the initial conditions as $t_0=1$ and $D(t_0)=1$. This equation provides an implicit solution for $D(t),$ which we must subsequently make explicit for each specific case.

For instance, let's consider both the sequences as power-law-like, i.e., $P_t = t^a$ and $N_D = t^b,$ where $a$ and $b$ are both greater than zero. These parameters are clearly related to the $\phi$ and $\psi$ of the sequences $\rho_t$ and $\nu_D.$ In this scenario, we have:
  \[
   \dfrac{dD}{D^b} \approx \dfrac{dt}{t^a} \qquad \implies \qquad \int_{1}^{D} \dfrac{dD}{D^b} \approx \int_{1}^t \dfrac{dt}{t^a}
  \]
  By solving these integrals and expressing the solution, we obtain:
  \[
   \dfrac{D^{1-b} - 1}{1-b} \approx \dfrac{t^{1-a}-1}{1-a} \qquad \implies \qquad
   D(t) = \left[ \dfrac{a-b}{a-1} + \dfrac{b-1}{a-1}t^{1-a} \right]^{\frac{1}{1-b}}.
  \] 
  The procedure remains the same for all other cases.
  
\section{Infinite Products}
\label{appendix_C}

In this appendix, we demonstrate that the probability, represented by the infinite product:
\[
   \mathcal{P}_{sat}=\prod_{t=T}^{\infty}\ton*{1-\frac{N_0+N_D}{P_t + N_D}}\approx\prod_{t=T}^{\infty}\ton*{1-\frac{\text{const}}{P_t}},
\]
converges to a finite nonzero number if and only if the sequence $P_t$ diverges at a rate of at least $\mathcal{O}(t)$.

To simplify the problem more broadly, let's consider a decreasing sequence $\{a_t\}$ with $t \in \mathbb{N}$. This sequence satisfies $a_t \to 0$ as $t \to \infty$ and ensures that $a_t < 1$ for all $t \in \mathbb{N}$. Our goal is to determine the conditions under which the infinite product:
\[
   \lim_{t \to \infty} \prod_{i=1}^{t} (1-a_t)=\prod_{t=1}^{\infty} (1-a_t) \equiv c
\]
converges to a finite number different from $0$.
  
First, it is important to notice that, since each term $(1-a_t)$ for all $t \in \mathbb{N}$ is less than $1$ but greater than $0$, the infinite product can either converge to a finite number strictly smaller than $1$ or converge to $0$. We are specifically interested in the conditions that lead to the product converging to a finite non-zero value, denoted as $c>0$.
  
To simplify the analysis, we take the logarithm of the infinite product, transforming it into an infinite summation:
\[
   \log \prod_{t=1}^{\infty} (1-a_t)=\sum_{t=1}^{\infty} \log(1-a_t) \equiv l \qquad \implies \qquad c=e^l
\]
and call its limit $l$. We can write the following conditions:
\begin{align}
   & \prod_{t=1}^{\infty} (1-a_t)=0 \qquad \iff \qquad \sum_{n=1}^{\infty} \log(1-a_t)=-\infty \nonumber \\
   & \prod_{t=1}^{\infty} (1-a_t)>0 \qquad \iff \qquad \sum_{t=1}^{\infty} \log(1-a_t) > -\infty \nonumber
  \end{align}
We now focus on understanding the behaviour of the sum of the logarithms. As $a_t \to 0$ for $t \to \infty$, after a certain point, we can approximate the logarithm as follows:
\[
   \sum_{t=1}^{\infty} \log(1-a_t) \sim \sum_{t=1}^{\infty} - a_t
\]
This results in a negative series that can either converge to a negative value or diverge to $-\infty$. According to the theorem of absolute convergence, the sum converges if and only if $\sum_{t=1}^{\infty} a_t$ converges, and it diverges if the series diverges. Thus, we have reduced the analysis to the behaviour of the elements in the sequence $a_t$. We can finally state:
\begin{align}
   & \prod_{t=1}^{\infty} (1-a_t)=0 \qquad \iff \qquad \sum_{t=1}^{\infty} a_t=\infty \nonumber \\
   & \prod_{t=1}^{\infty} (1-a_t)>0 \qquad \iff \qquad \sum_{t=1}^{\infty} a_t < \infty \nonumber
\end{align}
So if $a_t$ approaches zero at a rate of $t^{-1}$ or slower, the infinite product converges to zero. However, if it converges faster, the infinite product converges to a finite non-zero number, denoted as $0 < c < 1$

Applying this result to our case, where $a_t = \frac{\text{const}}{P_t}$, we can conclude that $\mathcal{P}_{sat}$ is greater than $0$ if and only if $P_t$ diverges at a rate of at least $\mathcal{O}(t)$ or faster.

\section{Detailed computations for the case $\phi=\psi$}
\label{appendix_D}

In this appendix, we provide a more detailed explanation of the case $\phi=\psi$ computations and explore examples of functional forms for $P_t$ and $N_D$ that differ from power laws.

We remind that, for $\phi = \psi$, $\nu_i$ and $\rho_i$ only differ by a constant that we have defined as $\mathcal{I}=\lim_{i\to\infty}\frac{\nu_i}{\rho_i}$. $\mathcal{I}$ quantifies the asymptotic relative importance of the reinforcement and triggering processes. Under these circumstances, one has that $N_D=\mathcal{I}P_D$, where $P_D$ is $P_t$ computed for the index $t=D$.

First, let's examine what happens when we substitute the form $D(t) = ct$, where $c$ is a constant in the interval $(0,1)$, into Eq.~\eqref{eq:D(t)_finiteI}. Assuming a power-law behaviour for $P_t \sim t^a$, we have:
\[
  c = \dfrac{\mathcal{I}(ct)^a}{t^a + \mathcal{I}(ct)^a} = \dfrac{\mathcal{I}c^a}{1+\mathcal{I}c^a}
\]
This can be further simplified to:
\[
  \mathcal{I}c^{a-1} = \mathcal{I}c^a + 1
\]
or equivalently:
\[
  c^{a-1}-c^a = \dfrac{1}{\mathcal{I}}
\]
To determine when this equation has a solution for some $c \in (0,1)$, we can analyse the left-hand side of the equation. The function $c^{a-1}-c^a$ has a singularity at $c=0$ when $a \in (0,1)$. This implies that the image of this function is the set of positive real numbers, ensuring a solution to the equation exists for every $\mathcal{I} > 0$. However, when $a>1$, the image of the function is limited in the domain $c \in (0,1)$, indicating that we cannot find a solution for every $\mathcal{I} > 0$. Therefore, the existence condition for the solution $D(t) = ct$ is $a \in (0,1)$.
  
To demonstrate that a saturating solution for $D(t)$ is not acceptable when $a<1$, we can use similar reasoning as in the case of $\phi<\psi$. If $D(t)$ saturates, Eq.~\eqref{eq:D(t)_finiteI} reduces to:
\[
  \dfrac{dD(t)}{dt} \approx \dfrac{\text{const}}{t^a}
\]
which leads to a solution of $D(t) \sim t^{1-a}$. When $a<1$, this solution diverges rather than saturates, leading to an inconsistency.
  
The saturating solution can exist only for $a>1$, and it is also the unique solution in this case, as $D(t) = ct$ is not acceptable.
  
Next, we check the existence of a sublinear regime by inserting the expression $D(t) = ct^k$ for some $k \in (0,1)$ into Eq.~\eqref{eq:D(t)_finiteI}. Following similar steps as before, we obtain the condition:
\[
  c k t^{k-1} \approx \mathcal{I} c^a t^{(k-1)a}
\]
This condition can be satisfied only when $a=1$ and $\mathcal{I}=k$, which corresponds to the case of the standard UMT~\cite{tria2014dynamics}.
  
Finally, we explore the cases of exponential or logarithmic sequences for $P_t$. In the case of exponential growth, Eq.~\eqref{eq:D(t)_finiteI} becomes:
\[
  \dfrac{dD(t)}{dt} \approx \dfrac{\mathcal{I}e^{D}}{e^t + \mathcal{I}e^{D}}
\]
It's evident that $D(t) = ct$ is not a solution in this case. However, a saturating solution is acceptable, leading to:
\[
  \dfrac{dD(t)}{dt} \approx \dfrac{\text{const}}{e^t} \quad \implies \quad D(t) \sim 1 - e^{-t}
\]
For the logarithmic case $P_t \sim \log t$, Eq.~\eqref{eq:D(t)_finiteI} becomes:
\[
  \dfrac{dD(t)}{dt} \approx \dfrac{\mathcal{I}\log D}{\log t + \mathcal{I}\log D}
\]
In this case, we cannot expect a saturating solution. However, if we substitute $D(t) = ct$, we obtain:
\[
  \dfrac{dD(t)}{dt} \approx \dfrac{\mathcal{I}\log ct}{\log t + \mathcal{I}\log ct} = \dfrac{\mathcal{I}\log t + \mathcal{I}\log c}{\log t + \mathcal{I}\log t + \mathcal{I}\log c}
\]
Neglecting the terms with $\log c$, we are left with:
\[
  \dfrac{dD(t)}{dt} = c \approx \dfrac{\mathcal{I}}{1+\mathcal{I}}
\]
which gives the value of the constant $c$.
  
\section{Examples of Size Distributions}
\label{appendix_E}

This Appendix provides practical examples of how to compute the rank-size distribution in the various scenarios considered in Fig.~\ref{fig:zipf}.
  
Let's consider the three regimes separately.

\begin{itemize}
  
\item $\boldsymbol{\phi>\psi}$: As derived in the main text, in this case, the total number of balls and the frequency of colour $i$ is given by:
\[
 N_i(t) \sim \rho_{t_i} \dfrac{P_t}{P_{t_i}} \qquad \qquad n_i(t) \sim \dfrac{t}{t_i}
\]
From these considerations, we can compute the complementary cumulative distribution, which is given by the formula:
\[
 p(n_i>n) = \dfrac{D\left( \dfrac{t}{n}\right)}{D(t)}
\]
Let's focus on the case when $\phi > -1$, thus $a = \phi + 1$. When $-1<\phi<0$, which means $0<a<1$, we observe Heaps' law in the number of distinct elements, $D(t) = t^{\gamma}$ for some $\gamma$. This gives a complementary distribution:
\[
 p(n_i > n) \approx n^{-\gamma}
\]
which is connected to a rank-size distribution 
\[
 f(R) = R^{-1/\gamma},
\]
recovering the usual relations between Heaps' and Zipf's exponents, known to be inversely related. Concerning the examples in Fig.~\ref{fig:zipf}a, calling $\alpha = 1/\gamma$ the Zipf exponent, we have that:
\begin{align}
 \gamma = 0.75 \qquad & \implies \qquad \alpha = 1.33 \nonumber \\
 \gamma = 0.5 \qquad & \implies \qquad \alpha = 2 \nonumber \\
 \gamma = 0.25 \qquad & \implies \qquad \alpha = 4 \nonumber
\end{align}
Then, if $\phi>0$, which means $a>1$, $D(t)$ saturates as $1-t^{-\phi}$. Substituting this, we obtain the complementary distribution:
\[
 p(n_i > n) \approx \dfrac{1-\left(\dfrac{t}{n} \right)^{-\phi} }{1-t^{-\phi}} \approx 1-\left(\dfrac{n}{t} \right)^{-\phi}
\]
This represents a flat complementary cumulative distribution, i.e., a power law with an exponent equal to $0$. Consequently, the rank-size distribution has an infinite exponent, indicating a step function. The saturation of $D(t)$ implies that at a certain time, no more new colours are introduced, and only the frequency of the already extracted elements keeps increasing in the dynamics. This generates a step function that drops after the rank corresponding to the last already extracted element; after this, no more elements will increase their frequency. 
   
These cases are represented by the last two curves in Fig.~\ref{fig:zipf}a. \\
   
\item $\boldsymbol{\phi<\psi}$: Let's first consider the case in which $-1 < \phi \leq 0$, where $D(t) \sim t$ with probability $1$. In this case, we can still easily compute the expression for $N_i(t)$ as in the example reported in the main text. Still, it is not straightforward to invert the relation to obtain the cumulative distribution $p(N_i>N)$ and $p(n_i>n)$.

However, we can deduce its form with a simple argument. $D(t) \sim t$ means that nearly every time a new element is extracted; thus no extractions can contribute to increasing the frequency of already extracted colours. In other words, each colour is hardly extracted more than once. This results in a step cumulative distribution (a power law with an infinite exponent) linked to a flat rank-size distribution (power law with exponent $0$).
   
This case is represented by the first two curves in Fig.~\ref{fig:zipf}b.
   
On the other hand, when $\phi>0$, we can observe both saturation and $D(t) \sim t$. When $D(t)$ saturates, we can still say:
\[
 N_i(t) \sim \rho_{t_i} \dfrac{P_t}{P_{t_i}} \qquad \qquad n_i(t) \sim \dfrac{t}{t_i}
\]
and
\[
 p(n_i>n) = \dfrac{D\left( \dfrac{t}{n}\right)}{D(t)}
\]
The reasoning in this case is similar to the case $\mathcal{I}=0$, resulting in a rank-size plot with an infinite steepness. 

This case is represented by the last two curves in Fig.~\ref{fig:zipf}b.\\

\item $\boldsymbol{\phi=\psi}$: We have to consider three regimes in this case. When $-1 < \phi < 0$, as computed in the main text, we have $D(t) = ct$ for some real constant $c \in (0,1)$. In this case, we have a complicated expression for $N_i(t)$:
\[
 N_i(t) \sim t_i ^{a-1}\left(\dfrac{t}{t_i}\right)^{\dfrac{a}{1+\mathcal{I}c^a}}
\]
from which it is not easy to derive an explicit form for $n_i(t)$. However, by analogy with the other cases, we can assume we can still use the general formula:
\[
 p(n_i>n) = \dfrac{D\left( \dfrac{t}{n}\right)}{D(t)}
\]
predicting a complementary cumulative distribution and a rank-size plot with an exponent of $-1$, which is well confirmed by the results of our simulations, as shown in Fig.~\ref{fig:zipf}c (blue and orange curves).
   
The other case is when $\phi = 0$; this represents the phenomenology of the standard UMT model. In this case, $\mathcal{I} = \dfrac{\nu}{\rho}$, and we have:
\[
 \begin{cases}
  D(t) = t^{\frac{\nu}{\rho}} \qquad \text{if } \nu < \rho \\
  D(t) = \dfrac{\nu-\rho}{\nu} t \qquad \text{if } \nu>\rho
 \end{cases}
\]
We obtain a rank-size distribution in both cases with an exponent of $\dfrac{\rho}{\nu}$. 
   
Finally, when $\phi > 0$, we have the usual:
\[
   p(n_i>n) = \dfrac{D\left( \dfrac{t}{n}\right)}{D(t)}
\]
Since $D(t)$ saturates, we have a step rank-size distribution (a power law with an infinite exponent). This case is represented in the last curve of Fig.~\ref{fig:zipf}c.
\end{itemize}

\section{Relation between the Hoppe's Urn model and TAP Equations}
\label{appendix_F}

The Hoppe's Urn model~\cite{hoppe1984polya} was introduced to describe the emergence of new alleles within gene sequences. The model was developed based on the observation that the probability that the $j$-th allele already occurred in a sequence was given by $j/(j+\theta)$ for a certain constant $\theta$. Hoppe introduced the following Urn model to represent this phenomenon abstractly.
  
Imagine an urn containing a black ball, which we call the replicator ball, and various other coloured balls. The black ball weighs $\theta$, while all the other balls weigh 1. Then, one randomly samples balls from the urn; when a ball of any colour is drawn, it is returned to the urn with an additional ball of the same colour. However, when the black ball is drawn, it is reinserted into the urn along with a ball of a new colour, symbolizing a triggering event. It's important to note that, in this formulation, the newly introduced colour is considered as observed. Consequently, the function $D(t)$ essentially counts the number of times the black ball has been extracted.

With this definition, the probability of extracting a new element at time $t$ is given by $\theta/(t+\theta)$, allowing us to express it as:
\[
  \dfrac{dD}{dt} = \dfrac{\theta}{t+\theta}
\]
This equation leads to logarithmic growth for $D(t)$, specifically $D(t) = \theta \log(t+\theta)$.
  
This model aligns seamlessly with the Time-dependent Urn Model with Triggering (TUMT) when we consider a constant reinforcement, $\rho_t = 1$, and either a null or decreasing adjacent possible. If, indeed, $\nu_D = 0$, and the initial size of the adjacent possible is $\theta$, or if $\nu_D$ is a decreasing function such that $N_D$ converges to a constant $\theta$, then we recover the Hoppe's Urn equation:
\[
  \dfrac{dD}{dt} \approx \dfrac{N_D}{P_t+N_D} = \dfrac{\theta}{t+\theta}
\]
Notably, this model works in the regime $\phi> \psi$. 
  
The TUMT can also be associated with the TAP equation~\cite{kauffman1996investigations} within the context of adjacent possible theory. TAP equations encompass a class of dynamical systems that explore the growth of different elements within a given space.
  
A standard form of a TAP Equation considers a system where elements grow proportionately to their combinatorial possibilities. In other words, at each time step, some possible recombination of elements occurs, generating a new element. Consequently, the dynamics of this system can be expressed as:
  \[
  M_{t+1} = M_t + \phi \sum_{i=1}^{M_t} \binom{M_t}{i}
  \]
The coefficient $\phi$ accounts for the recombination fraction at each time step. In the continuous-time approximation, we can simplify this expression as:
\[
  \dfrac{dM(t)}{dt} \approx \phi \sum_{i=1}^{M(t)} \binom{M(t)}{i} \approx \phi 2^{M_t}
\]
In this scenario, everything goes as if we were extracting a new element at every step, subsequently triggering the introduction of other elements into the urn (in proportion to the recombinations). Here, $D(t) = t$, and we can interpret $M(t)$ as representing the size of the adjacent possible, denoting the number of distinct elements in the urn not yet observed. Consequently, $M(t)$ corresponds to what we've referred to as $N_D = N_t$. 
  
The analogy between the TUMT and TAP equations holds when we set $\nu_D = \nu_t = \phi 2^{N_t} = \phi 2^{M(t)}$. As for the reinforcement, we can use whatever sequence $\rho_t$ such that $\lim_{i \to \infty} \dfrac{\nu_i}{\rho_i} = \infty$, in such a way that Eq.~\eqref{eq:D(t)_general} becomes:
\[
   \dfrac{dD(t)}{dt} \approx \dfrac{N_D}{P_t+N_D} \approx \dfrac{N_D}{N_D} = 1 \qquad \implies \qquad D(t) = t
\]
Indeed, the TAP equations are working in the regime $\phi<\psi$. The equation for $N_t = M(t)$ can then be expressed as:
\[
 \dfrac{dN_t}{dt} = \nu_t \dfrac{dD}{dt} = \nu_t \cdot 1 = \phi 2^{M(t)} = \dfrac{dM(t)}{dt}
\]
This result reinstates the analogy, and various forms of TAP equations can also be obtained by varying the specific expression for $\nu_t$.

\end{document}